\documentclass[twocolumn,amsmath,amssymb,floatfix]{revtex4}

\usepackage{bm}
\usepackage{amsmath}
\usepackage{epsf}
\def\simlt{\lesssim}
\def\simgt{\gtrsim}

\newcommand{\ApJ}{Astrophys. J}
\newcommand{\PRL}{Phys. Rev. Lett.}
\newcommand{\PRD}{Phys. Rev. D}
\newcommand{\MNRAS}{Mon. Not. R. Astron. Soc.}

\newcommand{\etal}{{\it et al. }}
\newcommand{\aut}[2]{{#2.\ #1,}}
\newcommand{\paut}[2]{{#2.\ #1} and}
\newcommand{\laut}[2]{{#2.\ #1,}}
\newcommand{\refs}[6]{#2, {\bf #3},  {#4} (#5).}
\newcommand{\mrefs}[6]{#2, {\bf #3},  {#4} (#5);}

\newcommand{\mybib}[2]{\bibitem{#2}}
\newcommand{\bn}{{\hat {\bf n}}}
\begin{document}

\title{Principal Power of the CMB}
\author{Wayne Hu$^{1}$ and Takemi Okamoto$^{2}$}
\affiliation{
{}$^{1}$Center for Cosmological Physics and 
Department of Astronomy and Astrophysics, 
University of Chicago, Chicago IL 60637\\
{}$^{2}$Department of Physics, 
University of Chicago, Chicago IL 60637\\
}

\begin{abstract}
We study the physical limitations placed on CMB 
temperature and polarization measurements of the
initial power spectrum by geometric projection, acoustic physics, 
gravitational lensing and the joint fitting of cosmological parameters.
Detailed information on the spectrum is greatly assisted by polarization information
and localized to the acoustic regime $k = 0.02-0.2$ Mpc$^{-1}$ with 
a fundamental resolution of $\Delta k/k>0.05$.  
From this study we construct principal component based statistics, which 
are orthogonal to cosmological parameters including the initial amplitude and
tilt of the spectrum, that best probe deviations from scale-free
initial conditions. 
These statistics resemble Fourier modes confined to the acoustic regime and
ultimately can yield $\sim 50$ independent measurements of the power spectrum
features to percent level precision.  They are straightforwardly related 
to more traditional parameterizations
such as the the running of the tilt and in the future can provide
many statistically independent measurements of it for consistency checks.
Though we mainly consider physical limitations on the measurements, these
techniques can be readily adapted to include instrumental limitations or
other sources of power spectrum information.
\end{abstract}
\maketitle

\section{Introduction}

High precision cosmic microwave background (CMB) temperature anisotropy 
measurements from WMAP have ushered in a new era in the determination
of the initial conditions for
structure formation and their implications for the physics 
of the early universe \cite{Peietal03}.  
Hints of deviations from power law initial
conditions from WMAP in the form of marginally significant coherent glitches 
over several multipoles and running of the tilt 
can be tested by the Planck satellite \cite{Planck}
and definitively measured with a high precision 
CMB polarization experiment such as the envisaged Inflation Probe
CMBPol \cite{CMBPol}.

Significant deviations would imply that the initial power 
spectrum is not scale free 
and hence traditional descriptions such 
as a constant running of the tilt, motivated by the small deviations 
expected from slow-roll inflation (e.g. \cite{MukFelBra92}),
 can lead to misinterpretation and
biases in parameter determination.

A large body of work already exists on more general approaches.  They
are typically
based on flat bandpowers, discrete sampling with interpolation 
for continuity or smoothness, wavelet expansions or more generally
some predetermined set of windows on, or effective
regularization of, the initial power spectrum
\cite{MukWan03,MilNicGenWas02,TegZal02,BriLewWelEfs03}.  

In this paper, we study the 
fundamental physical limitations placed on the measurement of
the initial power spectrum by geometric projection, acoustic physics,
gravitational lensing, and the joint fitting of cosmological parameters.
We then use this study to develop complementary techniques 
that are optimized with respect to the physics
of the CMB rather than prior assumptions about the form of the 
initial power spectrum.

The outline of this paper is as follows.  In \S \ref{sec:transfer}, we 
study the processes that transfer
fluctuations from the initial curvature perturbation to the observed temperature
and polarization fields and quantify the limitations imposed by acoustic
physics and geometric projection.  
In \S \ref{sec:lensing}, we introduce a discrete parameterization of
the initial spectrum that is matched to the further limitations imposed
by gravitational lensing.  In \S \ref{sec:fisher}, we employ Fisher matrix
techniques to study the effect of joint estimation of the initial spectrum
and cosmological parameters.  In \S \ref{sec:principal}, we construct
modes which best constrain 
from scale-free initial conditions out of its principal components.  
We conclude in
\S \ref{sec:discussion}.

\begin{figure}[tb]
\centerline{\epsfxsize=3.35truein\epsffile{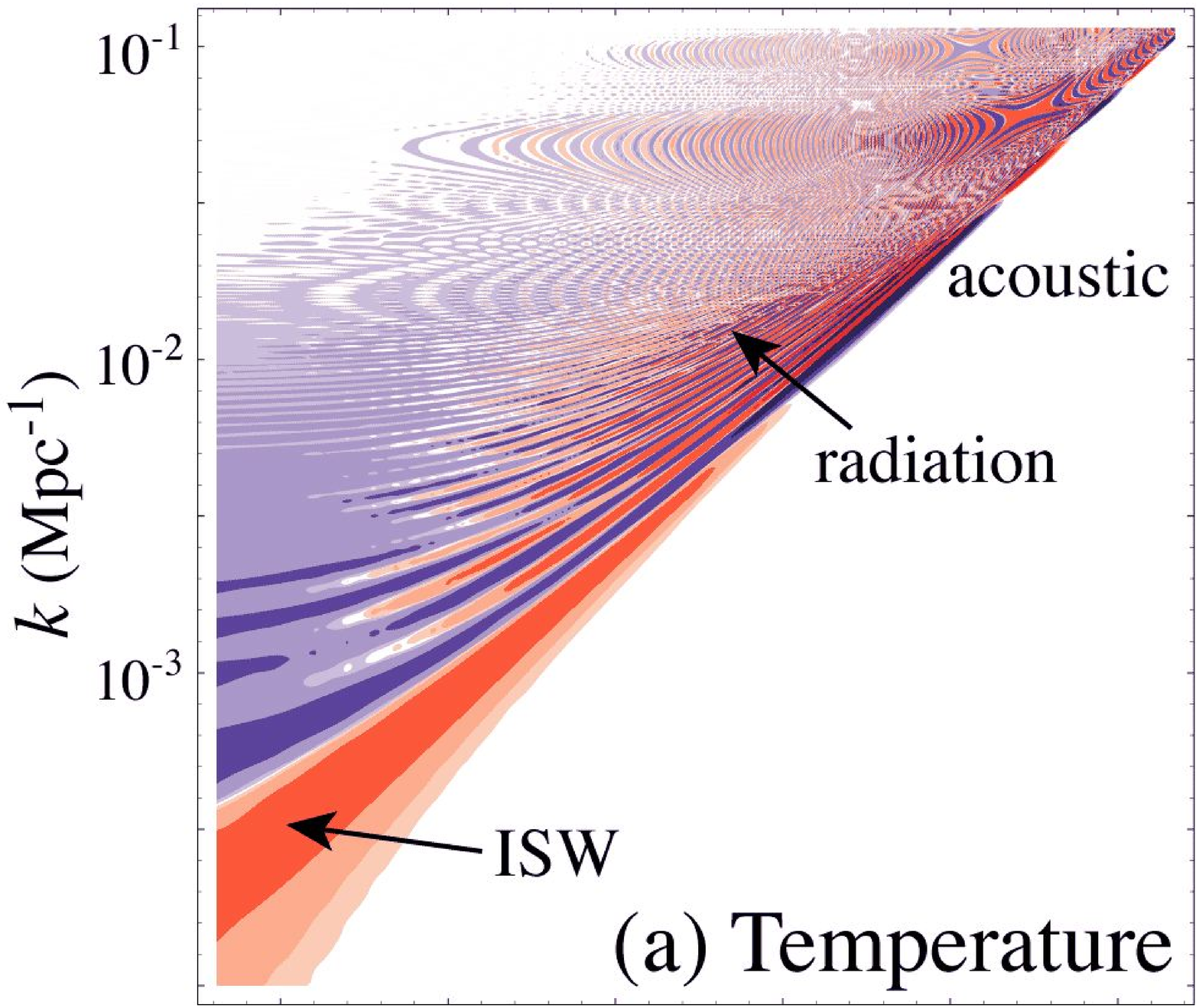}}

\centerline{\epsfxsize=3.35truein\epsffile{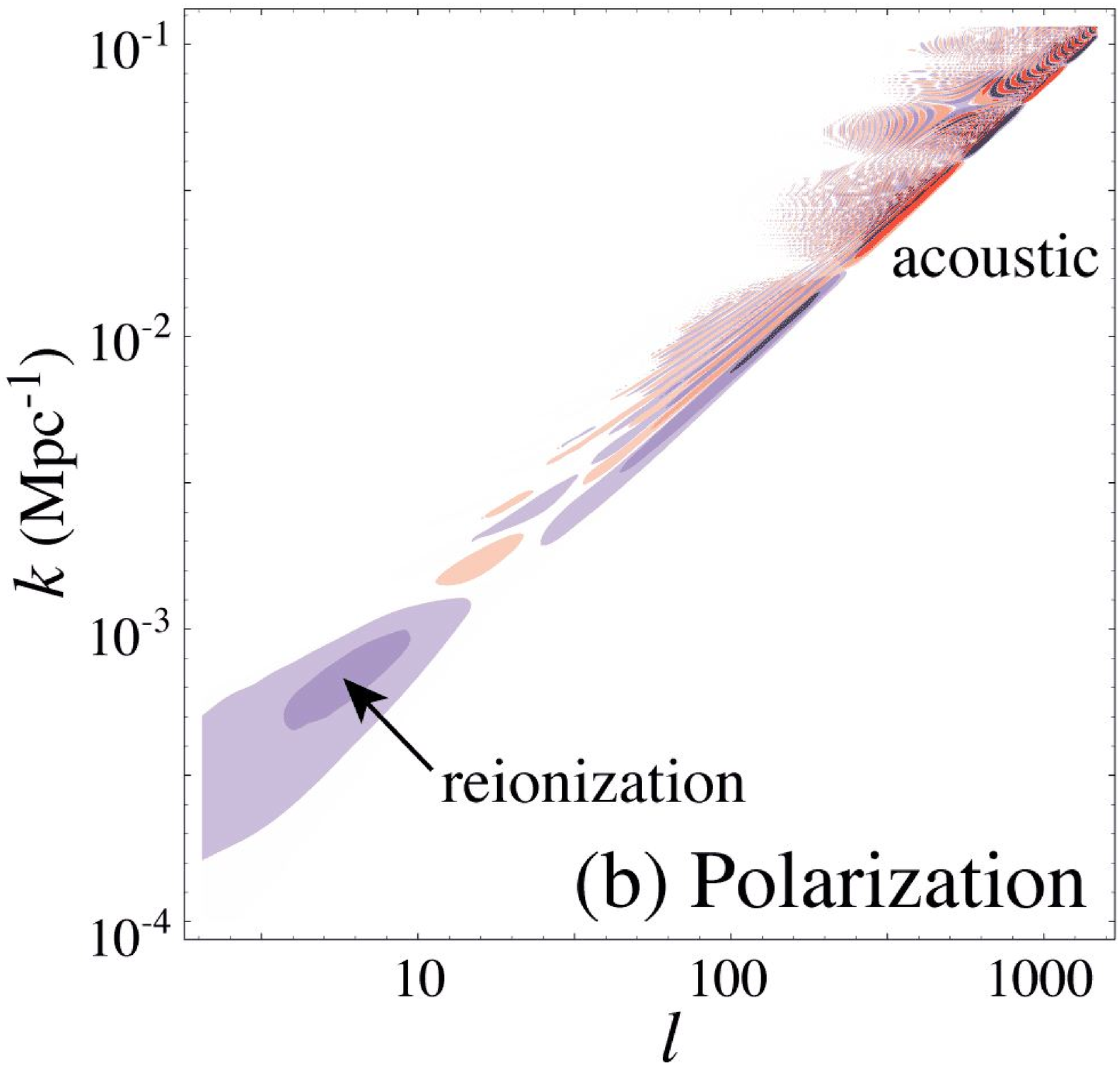}}
\caption{Transfer functions in temperature and polarization 
Contours are spaced by factors of 4 with blue (red) denoting fluctuations with
the same (opposite) sign as $\zeta$.
Note the projection ridge along the diagonal, the out-of-phase
temperature and polarization acoustic oscillations along the ridge 
beginning with the Sachs-Wolfe (SW) effect
for the temperature, the reionization bump for the polarization, and the
potential decay terms from the dark energy (ISW) and radiation 
for the temperature.  
The polarization transfer
is cleaner than and complementary to the temperature transfer.
}
\label{fig:transfer}
\end{figure}

\section{Transfer Functions and Projection}
\label{sec:transfer}

Acoustic and gravitational physics generate CMB temperature fluctuations
$\Theta=\Delta T(\bn)/T$ and polarization
fluctuations $E(\bn)$ \cite{ZalSel97} in the angular direction $\bn$ 
from an initial spatial curvature fluctuation $\zeta({\bf x})$, which we will
specify in the comoving gauge.
The observable angular power spectra are defined by the two point function of
their multipole moments
\begin{equation}
\langle X_{\ell m}^* X'_{\ell' m'} \rangle = \delta_{\ell\ell'}\delta_{mm'} 
	C_\ell^{XX'}\,,
\label{eqn:obspowerdef}
\end{equation}
where $X,X' \in \Theta,E$ and consequently are 
related to the initial curvature power spectrum
\begin{equation}
\langle\zeta^*({\bf k})\zeta({\bf k}')\rangle = (2\pi)^3 \delta({\bf k}-{\bf k}') 
	{2\pi^2 \over k^3}{\Delta^2_\zeta}(k)\,.
\label{eqn:initpowerdef}
\end{equation}
For slow-roll inflation the initial spectrum is given by (e.g. see \cite{MukFelBra92})
\begin{equation}
\Delta_\zeta^2(k) = {G \over \pi \epsilon} H^2_i\,,
\end{equation}
where $H_i$ is the Hubble parameter, $\epsilon = 3(1+w_i)/2$, and $w_i$ is the 
equation of state, all evaluated when the $k$-mode exited
the horizon during inflation.  
To the extent that $H_i$ and $w_i$ are constant, the spectrum
is scale invariant; to the extent that the evolution is small
across observable scales, the spectrum is a scale-free power law. 

The mappings between the two sets of power spectra are specified 
by transfer functions
\begin{equation}
{\ell (\ell+1) C_\ell^{XX'} \over 2\pi} = \int {d \ln k} \,
T^{X}_\ell(k) T^{X'}_\ell(k)\,
\Delta_\zeta^2(k)\,.
\label{eqn:transdef}
\end{equation}
Note that unlike the familiar transfer function of the matter power
spectrum, the CMB transfer functions are inherently two-dimensional since
they convert spatial fluctuations to angular fluctuations \cite{HuSug95b}.

These transfer functions are obtained by numerically
solving the linearized Einstein-Boltzmann equations.  
The publically available 
CMBFast \cite{SelZal96} and CAMB \cite{LewChaLas00} integral codes 
do return these functions in principle.  
However they implicitly assume that the initial
conditions are smooth so that the observable power spectra can be
rapidly calculated by interpolation in $\ell$ and $k$.
This interpolation can cause problems if the initial power spectrum
contains features that are comparable to the gridding, e.g. typically
$\Delta \ell=50$.
Traditional Boltzmann hierarchy codes 
\cite{WilSil81,BonEfs84,WhiSco96}
on the other hand solve the equations for every $\ell<\ell_{\rm max}$ but 
typically sample sparsely in the computationally costly 
$k$-modes and so require explicit smoothing in
$\ell$ \cite{HuScoSugWhi95}.  

We have instead developed an independent, fast version of a
Boltzmann hierarchy code whose speed without smoothing is comparable
to the public codes with an evaluation at every $\ell$.
Specifically the code solves the linearized equations 
\cite{HuSelWhiZal98} and the multilevel-atom calibrated 2-level 
hydrogen and helium recombination 
equations \cite{SeaSasSco99} in the comoving gauge 
out to $\ell_{\rm max}=6000$
with absorptive boundary conditions.
We have checked that the agreement with CMBFast v4.3 with RECFast
for smooth
initial conditions is typically at the few $10^{-3}$ level
and comparable to the numerical error
in CMBFast \cite{SelSugWhiZal03}.

\begin{figure}[tb]
\centerline{\epsfxsize=3.25truein\epsffile{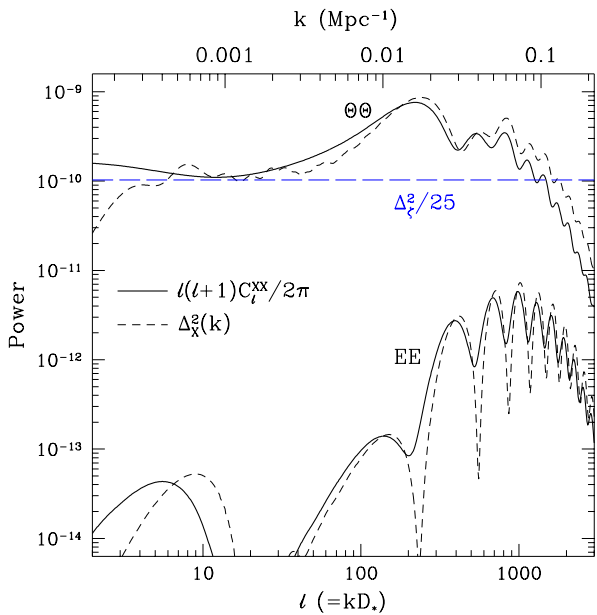}}
\caption{CMB angular and spatial power spectra for the fiducial cosmology
and spectrum $\Delta_\zeta^2 = (5.07 \times 10^{-5})^2$.  The spatial power
spectra $\Delta^2_X(k)$ are plotted with
$k=\ell/D_*$.  The low $k$ cutoff is due to $\ell_{\rm min}=2$ and
corresponds to superhorizon scale modes.  
Asymmetry in the projection leads 
to $\ell < k D_*$ power that is much more pronounced in the temperature 
than the polarization.}
\label{fig:cl}
\end{figure}

For this work, we oversample the transfer functions
with 3750 logarithmically spaced $k$-modes
from $k=10^{-4.35} - 10^{-1.35}$
Mpc$^{-1}$ ($\delta k/k = 1.8 \times 10^{-3}$) and 2500 $k$-modes from 
$10^{-1.35} - 10^{-0.35}$ Mpc$^{-1}$ ($\delta k/k = 0.9 \times 10^{-3}$).  
Figure~\ref{fig:transfer} shows the transfer functions for the fiducial
flat cosmology with dark energy density $\Omega_{\rm DE}=0.72$, dark energy
equation of state $w=-1$, matter density $\Omega_m h^2=0.145$, 
baryon density $\Omega_b h^2=0.024$ and reionization optical depth $\tau=0.17$.

The most notable feature of the transfer functions is the main projection ridge
along the diagonal at $k D_* = \ell$, where the angular diameter distance
to recombination $D_* = 13.8$Gpc in the fiducial model. 
It is instructive to consider first the simple Sachs-Wolfe (SW) 
\cite{SacWol67} limit of purely gravitational effects
in a flat universe. Here the temperature fluctuation 
is a projection of $1/3$ of the Newtonian gravitational potential or
equivalently $1/5$ of the comoving curvature,
\begin{eqnarray}
\Theta_{\ell m} &=& \int d\bn Y_{\ell m}^*(\bn) \Theta(\bn) \nonumber\\
%&= \int d\bn
%	Y_{\ell m}^*(\bn) {1 \over 3}\Psi({\bf x}=D_*\bn,D_*) \\
&=& \int d\bn
	Y_{\ell m}^*(\bn) {1 \over 5}\zeta({\bf x}=D_*\bn) \nonumber\\
&=& {4 \over 5}\pi \int {d^3 k \over (2\pi)^3} i^{\ell} j_\ell(kD_*) Y_{\ell m}^*
(\hat{\bf k}) \zeta({\bf k})\,,\nonumber\\
E_{\ell m} &=& 0\,.
\end{eqnarray}
Terefore comparing Eqn.~(\ref{eqn:obspowerdef})
(\ref{eqn:initpowerdef}) and (\ref{eqn:transdef}), we obtain
\begin{eqnarray}
T_\ell^\Theta(k) &=&{ \sqrt{2\ell(\ell+1)} \over 5} j_\ell(k D_*)\,, \nonumber\\
T_\ell^E(k) &=& 0\,.
\end{eqnarray}
By further assuming a fiducial model with a scale-invariant initial 
power spectrum
\begin{equation} 
\Delta_\zeta^2(k) \big|_{\rm fid} = \delta_\zeta^2 = (5.07\times 10^{-5})^2\,,
\end{equation}
we can evaluate the integral over $\ln k$ to obtain
\begin{eqnarray}
{\ell (\ell+1) \over 2\pi}C_\ell^{\Theta\Theta} &=& {\Delta_\zeta^2 \over 25}\,,
\nonumber\\
{\ell (\ell+1) \over 2\pi}C_\ell^{EE} &=& 0\,.
\end{eqnarray}
The sums over $\ln \ell$ yield the spatial power spectra
\begin{eqnarray}
\Delta_{\Theta}^2(k) &=& \sum_{\ell=2}^{\ell_{\rm max}} {1 \over \ell}
T_\ell^\Theta(k) T_\ell^\Theta(k) \Delta_\zeta^2 %\nonumber\\
%	&\approx& 
\approx {\Delta_\zeta^2 \over 25}\,,\quad (\ell\gg 1)\,, \nonumber\\
\Delta_{E}^2(k) &=& \sum_{\ell=2}^{\ell_{\rm max}} {1 \over \ell}
T_\ell^E(k) T_\ell^E(k) \Delta_\zeta^2 = 0\,,
\end{eqnarray}
which quantifies the contribution per log interval to the field variance 
$\langle X(\bn)X(\bn) \rangle$ \cite{BonEfs87}.  
Thus geometric projection represents
a power conserving mapping described by the 
spherical Bessel function $j_\ell(k D_*)$.   Under the
SW approximation the transfer function of Fig.~\ref{fig:transfer} would
appear as a diagonal band that narrowed with increasing $k$.

\begin{figure}[tb]
\centerline{\epsfxsize=3.25truein\epsffile{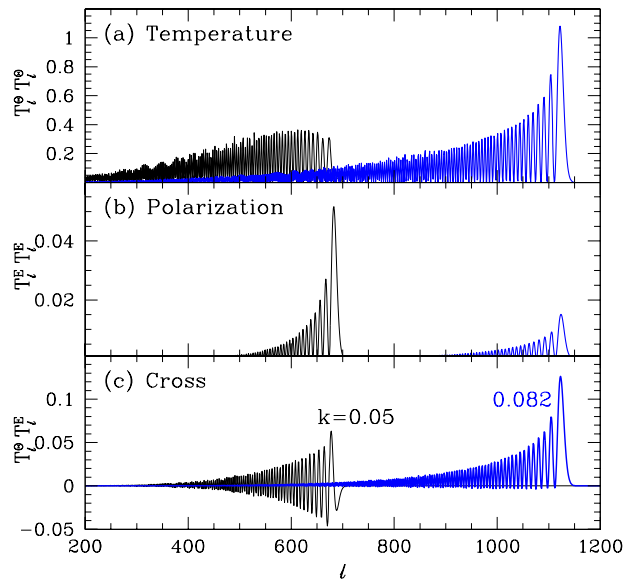}}
\caption{Angular spectrum of an individual $k$ mode for 
a trough $k_{m\approx 5/2}=0.05$ Mpc$^{-1}$ and a peak
$k_{m\approx 4}=0.082$ Mpc$^{-1}$.  At acoustic
troughs, the temperature perturbations are dominated by the Doppler
effect which is smeared out in angle by the projection; here the polarization
is at its maximum and yields a sharp projection of power.
}
\label{fig:kmode}
\end{figure}

The actual transfer functions differ from the SW approximation in interesting
ways.  Figure \ref{fig:cl} compares the $\ell$-space and $k$-space power 
spectra for the fiducial model.  The oscillations at high $k$ or $\ell$
are the familiar acoustic oscillations.  
For the temperature transfer function, the oscillations mainly trace 
the local plasma temperature at recombination and modulate the transfer
as $-\cos(k s_*)$.  
Here $s_*$ is the sound horizon at recombination;
$s_* = 144.5$ Mpc$^{-1}$ in the fiducial model.  We will number
the $m$th acoustic temperature peak as having wavenumber $k_m = m\pi/s_*$.

Figure \ref{fig:kmode} shows a mode near the $k_{m=4}$.
Though the projection has a peak at $\ell \approx kD_*$
with a narrow width of $\Delta \ell \sim 10-20$,
there is a strong asymmetry in the tails that skews the transfer  of
broadband power 
to $\ell < k D_*$.
This asymmetry reflects the fact that a wave that is nearly parallel to
the line-of-sight will have a projected wavenumber $\ell \ll k D_*$
as the $j_\ell(kD_*)$ factor in the transfer implies.
The asymmetry in the projection places the angular peaks
$\ell_m$ at values that are a few percent less than $k_m D_*$ \cite{HuWhi97a}.   
The asymmetry has a larger affect on the broadband power 
on small scales.  
Since photon diffusion during recombination
cuts off the spectrum at $k=k_D \approx 0.1$ Mpc$^{-1}$, 
a given $\ell$ in the damping tail
shows a deficit of broadband power from the oscillating tails of high $k$
modes.

Even in $k$-space, there is finite power at half integral $m$.
Here the plasma velocity induces a Doppler effect that modulates
the transfer function as $-\sin(k s_*)$.  
Moreover the $\ell$-structure of the transfer function
differs qualitatively from the SW result.  
Since the velocity transverse
to the line of sight yields no Doppler effect, the transfer function 
distributes $k$-power as $j_\ell'$ instead of $j_\ell$ \cite{HuSug95b}.  
In
Fig.~\ref{fig:kmode} we show a mode with $m=5/2$.  Notice that
there is no concentration of power near $\ell = k D_*$.   
Since power in these modes is distributed only broadband in $\ell$, recovery
of features in the initial power spectrum at these wavenumbers
from the temperature power spectrum will be severely limited.

\begin{figure}[tb]
\centerline{\epsfxsize=3.25truein\epsffile{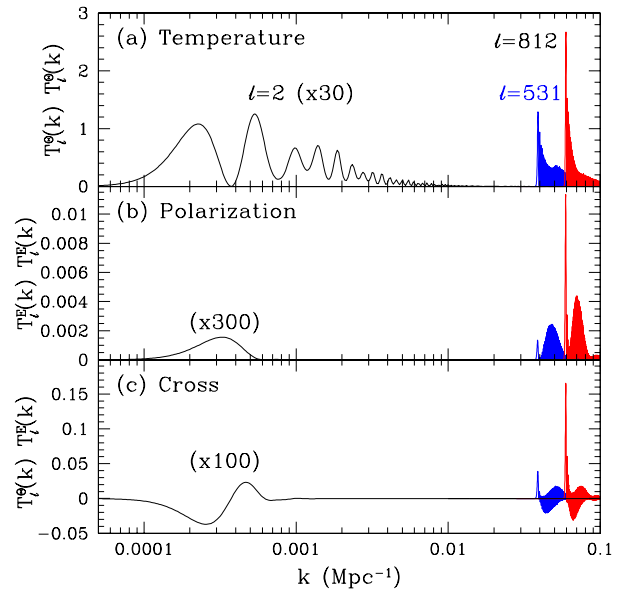}}
\caption{Spatial spectrum of an individual $\ell$ mode for 
a scale invariant initial spectrum.  The temperature quadrupole $\ell=2$ 
receives ISW contributions across
a wide range in $k$ while the polarization quadrupole has a much tighter
distribution.
At the acoustic peaks ($\ell_{m=2}=531$, $\ell_{m=3}=812$) the temperature
distribution is tighter but still receives contributions around 
$\sim k_{m+1/2}$
due to the broad Doppler distribution.  This effect is more prominent in the
polarization where the peaks represent deep minima that are filled in by
the projection broadened peaks.}
\label{fig:lmode}
\end{figure}

At $k \simlt \pi/s_*$, effects between recombination and the present
further broaden the transfer of power.
As the photons traverse decaying potentials wells (hills) 
during the end of radiation
domination and the onset of dark energy domination, they suffer a net
gravitational blueshift (redshift).   We call 
latter
the integrated Sachs Wolfe effect \cite{SacWol67} and the former
the radiation
effect (also known as the early ISW effect \cite{HuSug95b}).
Since the distance to the
observer is now $D<D_*$, the projection takes the power to lower $\ell$
or larger angles.   Conversely a single $\ell$ acquires power from
a broad range in $k$.  We show individual $\ell$-elements of the transfer function in 
Fig.~\ref{fig:lmode} for $\ell=2$, $\ell_{m=2}=531$ and $\ell_{m=3}=812$.  
Note the broad distribution in $\ln k$ for contributions to the
temperature quadrupole $\ell=2$.
Even at the $m$th peak in $\ell$ there are secondary contributions from
the wavenumbers near $k_{m+1/2}$ 
due to the broad projection of the Doppler effect.

The acoustic and projection limitations of 
the temperature power spectrum are largely removed
with polarization information.
Near the end of recombination,
photon diffusion creates local quadrupolar temperature anisotropy.  
Thomson scattering of radiation with a quadrupolar temperature anisotropy
linearly polarizes it.
The acoustic polarization transfer is therefore modulated according
to velocity gradients or $(k/k_D)\sin(k s_*)$ and like the Doppler effect 
is out of phase with the
local temperature.  
Unlike the Doppler effect, the projection of polarization
is sharp (see Fig.~\ref{fig:kmode}).  The 
spherical decomposition of plane wave $E$-polarization
projects as \cite{ZalSel97}
\begin{equation}
j_\ell(kD_*) \rightarrow \sqrt{{3 \over 8}{(\ell+2)! \over (\ell-2)!}} 
{j_\ell(kD_*) \over {(kD_*)^2}}\,,
\end{equation}
leading to a nearly diagonal transfer function (see Fig.~\ref{fig:transfer}).
Furthermore there is only one
source of acoustic polarization unlike the temperature
and so the power spectrum in 
$k$ space $\Delta^2_E(k)$ possesses zeros at $k_m$, where $m$ is an integer.
These sharp features in $k$ are only moderately smoothed by the projection.
Finally the low $k$, low $\ell$ end of the polarization transfer function 
is dominated by rescattering during reionization.  The projection remains
fairly sharp in comparison to the ISW dominated temperature projection 
though the shorter distance to the reionization epoch leads to a slightly
larger $\ell$ for a given $k$.

In summary, the temperature and polarization fields are complementary
in the transfer function, both in the out-of-phase acoustic power in $k$ 
and in the sharpness of the geometric projection to $\ell$.   
With both spectra and the cross correlation, there is 
high sensitivity to the initial power spectrum
in the whole acoustic range $k \sim 0.02-0.2$ Mpc$^{-1}$ with projection
limiting the sharpness of features in $\ell$ to 
$\Delta \ell \simgt 10-20$.  The reionization
bump in the polarization also offers a cleaner probe of the long wavelength
modes.  Although it is still severely limited by cosmic variance and the
determination of the reionization history from smaller scale modes,
polarization can in principle 
provide an incisive test for long wavelength power 
in our horizon volume. 
It can then determine whether the anomalously small low order temperature
multipoles found by COBE and WMAP 
are due to a lack of long wavelength power in our horizon volume
or a chance cancellation with more local effects.

\section{Discrete Representation and Gravitational Lensing}
\label{sec:lensing}

Due to the geometric projection and, as we shall see, gravitational lensing,
sharp features in the initial power spectrum will not be preserved in the
observable angular power spectra.  
For these reasons the transfer function
cannot be inverted directly to obtain the initial power spectrum
from the measured angular power spectra even with perfect measurements.   
Furthermore, as we shall
see in the next section, the transfer function itself depends on 
cosmological parameters that must be jointly estimated from the data.

Given the physical limitations of the transfer function it is 
convenient to represent the initial power spectrum by a discrete set
of parameters.
This discretization will also facilitate the treatment
of gravitational lensing below.  
Specifically, we represent the initial power spectrum as
\begin{equation}
\ln \Delta_\zeta^2(k) = \ln \delta_\zeta^2 + 
\sum_i p_i W_i(\ln k)\,,
\end{equation}
where $W$ is the triangle window
\begin{eqnarray}
W_i(\ln k) & = & {\rm max}\left(1- \big|{\ln k -\ln k_i \over \Delta \ln k}\big|,0\right) \,,
\label{eqn:window}
\end{eqnarray}
and we have assumed  a constant
\begin{equation}
\Delta \ln k=
\ln k_{i+1} -\ln k_i \,.
\end{equation}
The parameters $p_i$ then represent perturbations about the fiducial
scale invariant model with $\delta_\zeta=5.07 \times 10^{-5}$.  
An arbitrary initial power spectrum is then represented as the linear
interpolation in a log-log plot of points sampled at $k_i$.  If the
initial power spectrum contains features below the sampling rate then
the points represent averages through the windows.  This parameterization
preserves the positive definiteness of the initial power spectrum at the
expense of making the initial power spectrum non-linear in the parameters.

\begin{figure}[tb]
\centerline{\epsfxsize=3.25truein\epsffile{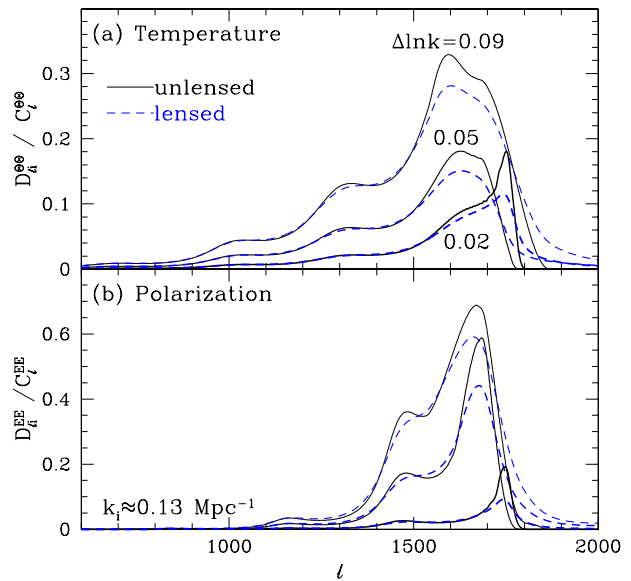}}
\caption{The power spectrum 
transfer matrix $D_{\ell i}^{XX}$ of a discrete
perturbation in the initial power spectrum at $\ln k_i$ of width $\Delta\ln k$.
Here $k_i \approx k_{m=6} \approx 0.13$ Mpc$^{-1}$. 
The transfer is broadened by projection and 
lensing such that only features of $\delta k/k > 0.05$ can be resolved.}
\label{fig:deriv}
\end{figure}

We can now define the infinitesimal response in the observable power spectra
to the initial power parameters 
\begin{eqnarray}
D_{\ell i}^{XX'} &=& {\partial C_\ell^{XX'} \over \partial p_i}\big|_{\rm fid} \label{eqn:derivdef}\\
&=& {2 \pi \over \ell(\ell+1)} \int d\ln k\,
T^{X}_\ell(k) T^{X'}_\ell(k)\, \delta_\zeta^2 W_i\,.
\nonumber
%\label{eqn:derivi}
\end{eqnarray}
This response function is a true transfer matrix for small perturbations $p_i$
from the fiducial model,
\begin{equation}
C_{\ell}^{XX'} \approx C_{\ell}^{XX'} \big|_{\rm fid} + \sum_i D_{\ell i}^{XX'}
p_i\,.
\end{equation}
In Fig.~\ref{fig:deriv} (solid lines) we show this transfer matrix for three choices of
$\Delta \ln k = 0.09$, $0.05$ and $0.02$ and $k_i \approx k_{m=6}
\approx 0.13$ Mpc$^{-1}$. Note that the intrinsic resolution from the
projection is only being reached at the finest binning for this high a $k$.  

The necessity of fine binning at high $k$ reflects a trade-off in the
choice of parameterization of the initial power spectrum.  
A logarithmic parameterization 
facilitates working with traditional approaches.  Inflationary
deviations from scale-invariance are usually quantified by 
the tilt $n$ and its running $\alpha$ 
\begin{eqnarray}
\ln \Delta_\zeta^2(k) &=& \ln \delta_\zeta^2 
+ (n-1) 
\ln({k\over k_{0}})
+ {1\over 2}\alpha \ln^2({k \over k_{0}})\,.
\end{eqnarray}
Thus the traditional parameterization is a Taylor expansion in log-log
space around the pivot point $\ln k_0$.  As a Taylor expansion, 
this parameterization can only be extended
to a large dynamic range $\Delta \ln(k/k_0) > 1$ 
if $\alpha \ll (1-n) \ll 1$ and
can yield misleading results if extended beyond its regime of applicability.
Nevertheless, the effect of $n$ and $\alpha$ can be easily constructed from the
discrete parameters
\begin{eqnarray}
p_i &\approx& (n-1) 
\ln({k_i \over k_{0}})
+ {1\over 2}\alpha \ln^2({k_i \over k_{0}})\,,
\label{eqn:tiltrunning}
\end{eqnarray}
to high accuracy with logarithmically spaced modes.

The drawback of a logarithmic representation is that the geometric projection 
causes a smoothing that is roughly linear in $\ell$ and hence the minimum
$\Delta \ell/\ell$ of features decreases with $\ell$.  However there is
a second fundamental limitation on features in the power spectrum that takes
over at high $\ell$: gravitational lensing \cite{Sel96b}.

Gravitational lensing deflects each 
line of sight according to the angular gradient in the projected potential $\phi$
\begin{eqnarray}
\bn &\rightarrow& \bn+\nabla \phi(\bn) \,,\nonumber\\
\phi(\bn) &=& 2 \int dD_c  {D_* - D \over D_* D} \Phi({\bf x}= \bn D,D_c)\,,
\label{eqn:projpot}
\end{eqnarray}
where the conformal lookback time $D_c=D$ in the flat fiducial model.
$\Phi$ is the Newtonian curvature and it is related
to the density perturbation in the comoving gauge by the Poisson equation.
In the fiducial cosmology,
an rms deflection of $\sigma_d \approx 3'$ 
arises from the projected potential
\begin{equation} 
\sigma_d^2 = \sum_\ell \ell(\ell+1) {2 \ell+1 \over 4\pi} C_\ell^{\phi\phi}
\end{equation}
and gets its main contributions from $\ell \sim 50$.
In other words, the arcminute scale deflections 
have a coherence of several degrees.  
Also important are the fluctuations at the arcminute scale which cause 
a deflection of a few arcseconds.

Following the linearized all-sky harmonic approach  \cite{Hu00b},
the change in the observed power spectrum is 
\begin{eqnarray}
\delta C_l^{XX'} &=& -[\ell(\ell+1)-{s_X^2+s_{X'}^2\over2}]
	{\sigma_d^2\over 2} C_\ell^{XX'} \\
\label{eqn:lens}
&&\quad +
	\sum_{\ell_1 \ell_2} C_{\ell_1}^{\phi\phi} C_{\ell_2}^{XX'}
	({}_{s_X} M_{\ell\ell_1\ell_2})
	({}_{s_{X'}} M_{\ell\ell_1\ell_2})\,, \nonumber
%\delta C_l^{EE} &=& -[\ell(\ell+1)-4] {\sigma_d^2 \over 2} C_\ell^{EE} +
%	\sum_{\ell_1 \ell_2} C_{\ell_1}^{\phi\phi} C_{\ell_2}^{EE}
%	{}_2 M_{\ell\ell_1\ell_2}^2 \nonumber\\
%\delta C_l^{\Theta E} &=& -[\ell(\ell+1)-2] {\sigma_d^2 \over 2} 
%	C_\ell^{\Theta E} \nonumber\\&&  +
%	\sum_{\ell_1 \ell_2} C_{\ell_1}^{\phi\phi} C_{\ell_2}^{\Theta E}
%	{}_0 M_{\ell\ell_1\ell_2}
%	{}_2 M_{\ell\ell_1\ell_2}  
\end{eqnarray}
where $s_\Theta=0$ and $s_E=2$ and the mode coupling term
\begin{eqnarray}
{}_s M_{\ell\ell_1\ell_2} &=&
{1\over 2} [\ell_1(\ell_1+1)+\ell_2(\ell_2+1)-\ell(\ell+1)]
\\&&\times
\sqrt{(2\ell_1+1)(2\ell_2+1) \over 4\pi } 
\left(
\begin{array}{ccc}
\ell & \ell_1 & \ell_2 \\
s & 0 & -s \\
\end{array}\right)\nonumber\,,
\end{eqnarray}
if $\ell+\ell_1+\ell_2 =$ even; 0 if odd.

This mode coupling leads to a broadening of features in the angular 
power spectra in two ways.
The main effect is a large-angle warping of the temperature and polarization 
maps that does not change the overall power in sub-degree scale
fluctuations.  
Here the two terms in Eqn.~(\ref{eqn:lens})
mainly cancel and the linearized approximation retains validity beyond
$\ell^2 \sigma_d^2/2 =1$ except for unphysical spectra with features
of $\delta \ell \sim 1$ \cite{Sel96b}.
However they induce a smoothing effect on features in 
the angular power spectrum and hence place a 
limit on resolving features in the initial curvature spectrum.  
In the absence of lensing, the resolution in harmonic space
$\Delta \ell$ is set by the fundamental mode of the survey, i.e. 
for the full sky $\Delta \ell=1$.
In the presence of lensing, the incoherence of the deflections in patches
subtending more than 
a few degrees effectively prevents their mosaicing to achieve higher
resolution.  This effect places a resolution limit of 
$\Delta \ell \sim 50$ for the
modes $\ell \simgt 10^3$ for which arcminute scale deflections are important.
Thus a sampling rate of $\Delta \ln k =0.05$ is sufficient even on these
scales. 

The second effect is from the smaller deflections with arcminute scale
coherence.   These induce additional arcminute scale fluctuations from
the distortion of primary fluctuations on larger scales \cite{Zal00}.
In the fiducial model, this secondary power becomes comparable to 
the primary power in temperature at $\ell_{\rm sec} 
\approx 3000$ and essentially eliminates the primary 
information on the initial power spectrum beyond $k \simgt \ell_{\rm sec}/D_*
\approx 0.02$ Mpc$^{-1}$.   Thus a sampling rate of $\Delta \ln k=0.05$ 
suffices for the whole observable regime.

On the other hand the lensing potential power spectrum $C_\ell^{\phi\phi}$
is also a source of information about the curvature fluctuations
at late times.  Unfortunately due to the projection in
Eqn.~(\ref{eqn:projpot}), lensing has
little ability to distinguish sharp features in the initial power spectrum
and even the interpretation of broadband power 
depends on the dark-energy and massive-neutrino dependent
growth rate of structure \cite{KapKnoSon03}. 
We ignore the information here by neglecting
the sensitivity to the parameters coming from $\partial C_\ell^{\phi\phi}/\partial
p_i$ and the information in the $B$-mode polarization.  
In this approximation, derivatives of the lensed CMB power spectra in
Eqn.~(\ref{eqn:derivdef}) are simply
given by the lensing of the power spectrum derivatives calculated from
the transfer functions.
In Fig.~\ref{fig:deriv} (dashed lines) we show the effect of gravitational
lensing on the power transfer.  Now even features with
$\Delta \ln k = 0.05$ are limited by lensing.  We will hereafter adopt this
spacing as the fiducial choice which leads to 200 parameters for the 
initial power spectrum.   

\begin{figure}[tb]
\centerline{\epsfxsize=3.25truein\epsffile{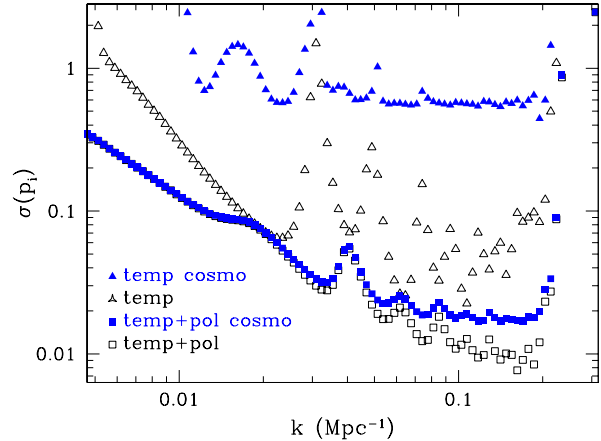}}
\caption{Individual errors on the discretized initial power spectrum $p_i$ with
$\Delta\ln k=0.05$ and a cosmic variance limited measurement out to
$\ell_{\rm sec}=3000$.  With fixed cosmological parameters, 
temperature information alone (open triangles) 
recovers $p_i$ for modes near the acoustic peaks 
but not the troughs.  Adding polarization information improves
coverage to the whole acoustic range $k=0.02-0.2$ Mpc$^{-1}$ 
(open squares).  Adding cosmological parameters destroys 
individual parameters constraints with just temperature information
(filled triangles) which are largely recovered by adding polarization
information (filled squares).  Much of the constraining power is
hidden in the strongly anticorrelated covariance (not shown).}
\label{fig:rawerr}
\end{figure}

\section{Statistical Forecasts and Cosmological Parameters}
\label{sec:fisher}

A third limitation on the recovery of the initial power spectrum comes
from degeneracy with cosmological parameters such as the energy density
of the particle constituents of the universe today.   
We employ the Fisher matrix technique
to study the effect of degeneracies.  Construction of the
Fisher matrix will also be the first step in building statistics
that are immune to cosmological parameter uncertainty.

We begin by appending the $n_c$ cosmological parameters to the
initial power vector $p_i$ to form the joint parameter vector 
$p_\mu$, $\mu = 1,\ldots,n_k,\ldots, n_k+n_c$.
The Fisher matrix for a cosmic variance limited 
measurement out to $\ell=\ell_{\rm sec}$ is
\begin{equation}
F_{\mu\nu} = 
\sum_{\ell=2}^{\ell_{\rm sec}}
{2 \ell +1 \over 2} 
{\rm Tr} [{\bf D}_{\ell \mu} 
{\bf C}_{\ell}^{-1}
{\bf D}_{\nu \ell} {\bf C}_{\ell}^{-1}]\,,
\label{eqn:fisher}
\end{equation}
where we have suppressed the $XX'$ indices in a matrix notation 
for the derivatives and the power; here 
\begin{equation}
({\bf D}_{\ell \mu})_{XX'} =  D_{\ell\mu}^{XX'} 
\equiv {\partial C_\ell \over \partial p_\mu}
\end{equation}
is the generalization of the transfer matrix Eqn.~(\ref{eqn:derivdef}).  
Given gravitational
lensing constraints and other sources of foreground
contamination, we will take $\ell_{\rm sec}=3000$.

The inverse Fisher matrix approximates the covariance matrix of 
the parameters.
In particular, the marginalized 1-$\sigma$ 
errors on a given parameter $\sigma(p_\mu)$ are given by
\begin{equation}
\sigma^2(p_\mu) = C_{\mu\mu} \approx ({\bf F}^{-1})_{\mu\mu}\,.
\end{equation}
The Fisher matrix also allows for a simple conversion between
representations of the parameter set $p_a$ and $p_\mu$
\begin{equation}
F_{\rm a b} = \sum_{\mu\nu}
	{\partial p_\mu \over \partial p_a} 
	F_{\mu\nu}
	{\partial p_\nu \over \partial p_b} \,,
\label{eqn:reparam}
\end{equation}	
where $p_a$ can have a lower dimensionality, e.g. by going from
the initial power parameterization of $p_i$ to $n$ and $\alpha$
through Eqn.~(\ref{eqn:tiltrunning}). 

In Fig.~\ref{fig:rawerr}, we show the statistical errors
on the initial power spectrum parameters $p_i$.
Even in the absence of cosmological parameters ($n_c=0$)
the constraints on the $p_i$ are strongly modulated by the
acoustic oscillations with temperature information alone.
The $k$-modes associated
with the acoustic troughs (half integral $m$) 
are poorly constrained while those
associated with the peaks (integral $m$) 
are fairly well constrained.  
For low $k$, both the projection and cosmic variance severely limits
the resolution in $k$ leading to a strong anticorrelation 
between points.  Hence only broadband power and not features can be recovered. 
For $k> 0.2$ Mpc$^{-1}$ the cut off at $\ell_{\rm sec}=3000$ limits the
information.  

With the addition of the polarization and cross spectra information,
the modulation of
errors with the peaks reverses.  Now modes associated with acoustic troughs
are better constrained than those associated with the peaks.  
The polarization information becomes even more important once the
cosmological parameters are added back in.  In Fig.~\ref{fig:rawerr}
we take 6 additional 
parameters $\Omega_b h^2$, $\Omega_m h^2$, $\Omega_{\rm DE}$, $w$, 
$\tau$ and $r$, where $r$ is the tensor-scalar ratio. 
With temperature information alone, the ability to measure an individual 
$p_i$ is essentially completely destroyed since the response to a cosmological
parameter can always be mimicked by a particular superposition of $p_i$'s 
\cite{Kin01}.  
However only six such combinations are lost and so
with the addition of polarization information with its complementary 
transfer function, the effect of marginalizing cosmological parameters
is fairly small.  In fact even with temperature information alone, there
are modes that are very well constrained.  The raw marginalized errors
on $p_i$ are misleading in that they do not show the covariance.   
It is therefore useful to find a more efficient representation of the 
information that accounts for projection, lensing and cosmological parameters.

\begin{figure}[tb]
\centerline{\epsfxsize=3.25truein\epsffile{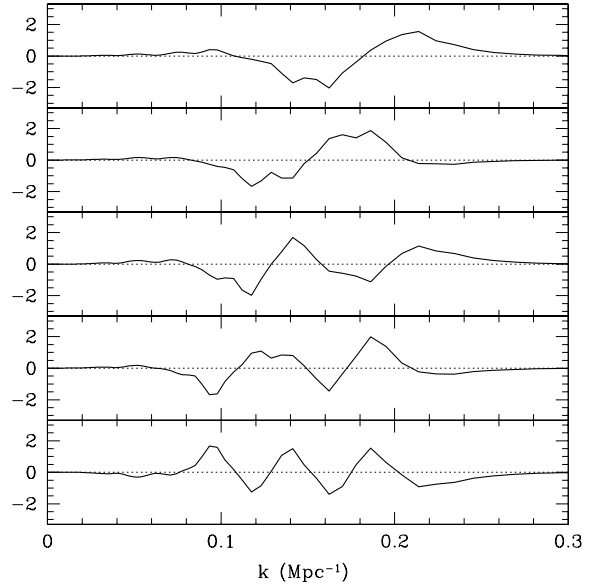}}
\caption{First 5 principal components of the covariance matrix with 6 
cosmological
parameters, the amplitude $\ln \delta_\zeta$ and tilt $n$ marginalized. 
The modes are normalized to unit variance over $d\ln k$ (discretized
to $\Delta\ln k=0.05$).  Well-constrained modes are 
localized in the acoustic regime 
$k=0.02-0.2$ Mpc$^{-1}$ and give zero weight to scale-free deviations.}
\label{fig:eigenvec}
\end{figure}

\section{Principal Components and Deviated Spectrum}
\label{sec:principal}

\begin{figure}[tb]
\centerline{\epsfxsize=3.25truein\epsffile{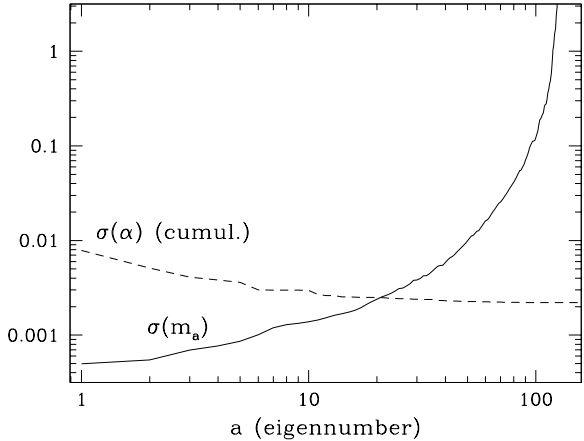}}
\caption{Eigenvalues and cumulative constraints on the running of the
tilt $\alpha$ represented in modes.  For cosmic variance limited
temperature and polarization measurements out to $\ell_{\rm sec}=3000$, 
approximately 50 modes can be measured to percent level precision ($\sigma(m_a)
< 0.01$). Information
on the running $\alpha$ saturates with more that $\sim 10$ modes.}
\label{fig:eigenval}
\end{figure}

Given the limitations imposed by projection, lensing and cosmology
what kinds of deviations from a scale-free initial spectrum
can the CMB best probe?
To answer this question, we now consider a principal component representation
of features.

The information retained after including these three effects is
quantified by the covariance matrix of the deviation parameters $p_i$, the
$C_{ij}$ subblock of the joint covariance matrix $C_{\mu\nu}$.
Employing the Fisher matrix approximation of the last section, we can
determine the principal component representation based
on the orthonormal eigenvectors $S_{i a}$ of $C_{ij}$
\begin{equation}
C_{ij} = A^{-2} \sum_{a} S_{i a} \sigma_{a}^{2} S_{j a}\,,
\end{equation}
where $A$ is a normalization parameter.
For a fixed $a$, $S_{i a}$ specifies a linear combination of
the deviation parameters $p_i$ for a new representation of the spectrum
\begin{equation}
m_a  =  A \sum_i S_{ia} p_i\,,
\end{equation}
where the covariance matrix of the mode amplitudes
is given by
\begin{equation}
\left< m_a m_b \right> = \sigma_a^2 \delta_{a b}\,.
\label{eqn:orthogonality}
\end{equation}
This relationship can also be derived from Eqn.~(\ref{eqn:reparam}).
In other words,
the eigenvectors form a new basis that is complete and yields uncorrelated
orthogonal measurements with variance
given by the eigenvalues.  We choose the normalization $A=(\Delta\ln k)^{-1/2}$
so that a $m_a=1$ corresponds to a deviation with 
unit variance when integrated over $d\ln k$, i.e.
\begin{equation}
A^2 \int {d\ln k}\, S_{i a}^2 \approx 
\sum_i S_{i a}^2 = 1\,.
\end{equation}
Modes corresponding to combinations that are degenerate with cosmological
parameters or due to the projection and lensing have high eigenvalues
and high variances and may be removed without loss of information.

In Fig.~{\ref{fig:eigenvec}}, we show the first five eigenvectors of the
covariance matrix including both temperature and polarization information.  
In this case we append $\ln \delta_\zeta$
(or $\ln A$) and $n$ to the list of cosmological parameters included 
in \S \ref{sec:fisher}.  By doing so, the best constrained eigenvectors 
automatically null out changes in the amplitude and tilt and so can be
interpreted as the best modes to search for deviations from scale-free
initial conditions.  They also tend to null out any slowly varying
component from residual secondary effects and foregrounds.
As a consequence, their form is similar to
Fourier modes confined to the well-measured acoustic regime $0.02-0.2$ Mpc$^{-1}$
with the constant and gradient term (in $\ln k$) removed.   
For the cosmic variance 
temperature and polarization spectra assumed in the Fisher matrix
calculation, the mode errors are shown in Fig.~\ref{fig:eigenval}
(solid line) and remain below the percent level for $\sim 50$ modes.

Note that truncation of the eigenmodes
does not limit types of initial spectra that can be tested against the
measurements $m_a$ since any spectrum may be projected onto the measurement
basis.  As an example, consider constraints on a constant 
running of the tilt $\alpha$ 
in the measurement basis.  Each mode independently measures $\alpha$ with
a variance
\begin{equation}
\sigma_a^2(\alpha) = A \sum S_{ia} {1\over 2} \ln^2({k_i \over k_0})\,,
\end{equation}
where the pivot $k_0$ is arbitrary since the $S_{ia}$ null out a constant deviation.
Since each measurement is independent, the total variance is given by
\begin{equation}
{1 \over \sigma^2(\alpha)} = \sum_{a=1}^{a_{\rm max}} {1 \over \sigma^2_a(\alpha)}\,.
\end{equation}
This cumulative sum is shown in Fig.~\ref{fig:eigenval} (dashed line).  
Information on $\alpha$ is efficiently captured in the first 10 eigenmodes.

\begin{figure}[tb]
\centerline{\epsfxsize=3.25truein\epsffile{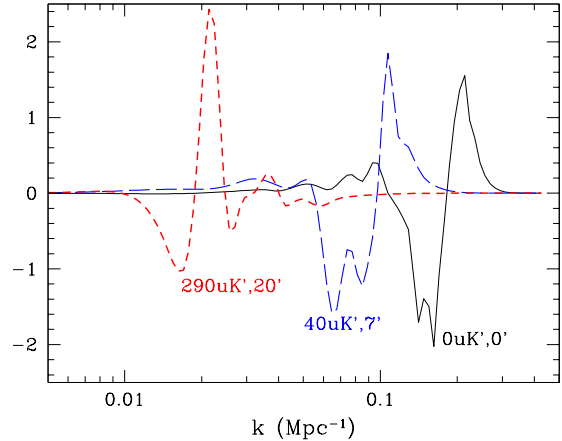}}
\caption{Eigenvectors in the presence of instrumental noise (in $\mu$K-arcmin)
and beam (FWHM in arcmin).  Parameters are chosen to roughly bracket the
expectations of WMAP and Planck (dashed lines) for comparison with the
noise-free case (solid lines).  The eigenvectors are now confined 
to $\ell$-modes and hence $k$-modes that are larger than the beam. 
In the highest noise, large beam case, polarization information is degraded 
and the avoidance of cosmological parameter degeneracies 
complicates the mode structure.
}
\label{fig:noise}
\end{figure}

The principal component technique can be made into a practical
tool for searches for deviations from scale-free 
initial curvature fluctuations.  
The best measured modes are fairly smooth and take forms that
are independent of the discretization $\Delta \ln k$ 
of the underlying representation. Hence they may be replaced by a 
continuous representation for parameter estimation with an efficient
integral code such as CMBFast or CAMB.    The parameters $m_a$ can
then be added in the usual way to a Monte Carlo Markov Chain.
Deviations of the true model from the assumed fiducial model in 
constructing the eigenmodes and non-linearities in the parameter
dependence will degrade the statistical orthogonality of modes
implied by Eqn.~(\ref{eqn:orthogonality}).  These covariances will
nonetheless be captured in the likelihood exploration.  The process
may also be iterated if the assumed eigenvectors deviate substantially
from those of the best fit cosmology.

Furthermore, although we
have assumed a perfect experiment here to determine the limitations imposed
by physical processes on feature recovery, the techniques are readily 
optimized for specific experiments.  One simply replaces the
power spectra in  Eqn.~(\ref{eqn:fisher}) 
for the Fisher matrix with the total signal and noise power spectra of
a given experiment.  
In Fig.~\ref{fig:noise}, we show a simple example where the experiment 
is parameterized by a white noise level in $\mu$K-radian $\Delta T$
and the FWHM of a
Gaussian beam $\theta$.  Here
\begin{eqnarray}
C_\ell^{\Theta\Theta}\big|_{\rm noise} &=& ({\Delta T \over T})^2 
e^{\ell(\ell+1) \theta^2/8\ln 2}\,,
\nonumber\\
C_\ell^{EE}\big|_{\rm noise} &=& 2({\Delta T \over T})^2 
e^{\ell(\ell+1) \theta^2/8\ln 2}\,,
\nonumber\\
C_\ell^{\Theta E}\big|_{\rm noise} &=& 0\,.
\end{eqnarray}
We show the best eigenmode for
two examples that roughly bracket the expectations for the
completed WMAP and Planck missions.
Note that the main effect is to replace the cutoff imposed by $\ell_{\rm sec}$
with that of the beam.  However the smaller $\ell$ range and the loss of
polarization in the high noise example complicate the form of the mode because
it adjusts to remain orthogonal to the cosmological parameters.

\section{Discussion}
\label{sec:discussion}

Physical limitations imposed by the formation and evolution of CMB temperature
and polarization anisotropy limit the ability to recover the initial
power spectrum from even perfect measurements of angular power spectra.  
The initial power spectrum
can be well measured in the acoustic regime $k=0.02-0.2$ Mpc$^{-1}$ but 
projection and gravitational lensing prevents resolution of features
smaller than $\Delta k/k \approx 0.05$.  The complementary acoustic
polarization information greatly assists in filling in the coverage
near the acoustic troughs and in immunizing the determination to
cosmological parameter degeneracies.  Furthermore polarization 
in principle offers
a cleaner probe of long-wavelength power through reionization effects.
Beyond purely two-point statistics in the temperature and polarization fields, 
the limitations of gravitational lensing can in principle be removed 
\cite{HuOka01,HirSel03}.   This can moderately improve the resolution of
features but will only be possible with ultra-high precision polarization
measurements across nearly the whole sky.

We have constructed the best statistics for 
measuring deviations from
scale-free initial conditions given these physical limitations on
power spectra.  They are constructed from the principal components of
the covariance matrix. 
These modes resemble a Fourier expansion of the
initial spectrum that is localized to the acoustic regime.  Approximately 
50 modes in this regime ultimately
can be measured with percent level precision. 
These modes can provide many independent measurements
of model-motivated parameters such as the running of the tilt as a 
consistency check that such simpler parameterizations are adequate.

Although beyond the scope of this work,
the principal component technique can be developed into a practical tool 
for the search for deviations from scale-free conditions
with current experiments.  These may also include non-CMB sources with
the appropriate addition of Fisher matrices.
Here the physical limitations imposed on the modes would be supplemented by 
experimental limitations.  
The principal component technique provides a general, physically-motivated
and complementary tool for the study of the initial power spectrum
and its implications for early universe physics.

{\it Acknowledgments:} 
This work was supported by NASA NAG5-10840 and
the DOE OJI program.  We thank 
the Aspen Center
for Physics and the participants of the CMB workshop, epecially endgamers 
S. Bridle, M.  Kaplinghat, A. Lewis, S. Myers, B. Wandelt, and Y. Wang.

\end{document}